\documentclass[aps,prl,twocolumn,groupedaddress,showpacs,superscriptaddress,amssymb,amsmath]{revtex4-1}
\usepackage{graphicx}
\usepackage{subfigure}
\usepackage{amsmath}
\usepackage{amssymb}
\usepackage{amsfonts}
\usepackage{color}
\usepackage{bm}        % for math
\usepackage{comment}
\usepackage{placeins}
\usepackage[mathscr]{eucal}
\usepackage[colorlinks, linkcolor=red,citecolor=blue,urlcolor=blue]{hyperref}
\usepackage[normalem]{ulem}
\usepackage[all]{hypcap}
\usepackage{multirow}
\usepackage[dvipsnames,usenames,table]{xcolor}
\usepackage{dcolumn}
\usepackage{hyperref}
\usepackage{bm}
\usepackage{epsf}
\usepackage{subfigure}
\usepackage{amsfonts}
\usepackage{epstopdf}%
\newcommand{\be}{\begin{equation}}
\newcommand{\ee}{\end{equation}}
\newcommand{\bea}{\begin{eqnarray}}
\newcommand{\eea}{\end{eqnarray}}
\def\langled{{\langle \langle}}
\def\rangled{{\rangle \rangle}}
\def\la{{\langle }}
\def\ra{{\rangle}}

\begin{document}
\title{Universal bounds on fluctuations for machines with broken time-reversal symmetry}
\author{Sushant Saryal}
\affiliation{Department of Physics,
		Indian Institute of Science Education and Research, Pune 411008, India}

\author{Sandipan Mohanta}
\affiliation{Department of Physics,
		Indian Institute of Science Education and Research, Pune 411008, India}

\author {Bijay Kumar Agarwalla}
\affiliation{Department of Physics,
		Indian Institute of Science Education and Research, Pune 411008, India}
		\email{bijay@iiserpune.ac.in}
		
\date{\today}

\begin{abstract}
For a generic class of machines with broken time-reversal symmetry we show that in the linear response regime the relative fluctuation of the sum of  output currents for time-forward and time-reversed processes is always lower bounded by the corresponding relative fluctuation of the sum of input currents. This bound is received when the same operating condition, for example, engine, refrigerator or pump, is imposed in both the forward and the reversed processes. As a consequence, universal upper and lower bounds for the ratio of fluctuations between the output and the input current is obtained. Furthermore, we establish an important connection between our results and the recently obtained generalized thermodynamic uncertainty relation for time-reversal symmetry broken systems.  We illustrate these findings for two different types of machines: (i) a steady-state three-terminal quantum thermoelectric setup in presence of an external magnetic field, and (ii) a periodically driven classical Brownian heat engine. 
\end{abstract}

\maketitle 

{\it Introduction.--}
Study of thermal fluctuations have played a significant role in shaping the field of statistical mechanics. For systems near thermal equilibrium, fluctuation-dissipation relation \cite{FDT1, FDT2, FDT3,fluc-diss, Maes} has emerged not only as a fundamental relation but also with practical implications giving a route to compute linear transport coefficients from equilibrium correlations \cite{GK1, GK2}. Since then significant efforts have been made to write down analogous relations for systems close to a non-equilibrium steady state \cite{ SSR1,SSR2,SSR-expt,SSR-latest}. Over the past two decades, discoveries of  various types of universal fluctuation relations \cite{fluc-1, fluc-2, fluc-3, st-thermo1, st-thermo2, st-thermo-Broeck, fluc1-Evans, fluc2-Evans, Jar,Q-thermo2} have further enhanced our understanding about thermodynamics of non-equilibrium systems beyond the typical linear response regime. In recent times, intense efforts have been directed towards understanding the impact of fluctuations on small scale non-equilibrium systems and more specifically on the performance of operational machines \cite{fluc-heat-1, fluc-heat-2, fluc-heat-3,Watanabe,Sagawa,bijay-engine1, bijay-engine2, Q-thermo2}.  In this regard, recently discovered thermodynamic uncertainty relations (TURs) \cite{Barato:2015:UncRel, trade-off-engine, Gingrich:2016:TUP,Falasco,Garrahan18,Timpanaro,Saito-TUR,Junjie-TUR, Agarwalla-TUR, Agarwalla-TUR0, Giacomo-slow, constancy,GooldPRR} have provided new and intriguing insights for autonomous and driven thermal machines by revealing a universal trade-off type relation involving efficiency, output power and its fluctuation \cite{trade-off-engine,constancy, Giacomo-slow}. 

Independent to TUR studies, there are recent interests in providing universal bounds for the ratio of higher order work and heat fluctuations for thermal machines \cite{Watanabe,Sagawa}. In this context, very recently, for finite-time four-stroke heat engines a universal upper bound was reported \cite{Watanabe}, given as  $ \eta^{(2)}  \equiv \la \la w^2 \ra \ra / \la \la Q^2 \ra \ra \leq \eta_C^2$ where $\eta_C = 1- T_c/T_h$ being the  Carnot efficiency and $T_h (T_c)$ is the temperature of the hot (cold) bath. Furthermore, it was shown in Ref.~\cite{bijay-universal} that for an autonomous time-reversal symmetric steady-state setup, operating as a useful machine (e.g., engine, refrigerator or pump), the relative fluctuation of output current is always lower bounded by the corresponding relative fluctuation of the input current.
%whenever the setup works as a useful machine. 
As an immediate consequence of this result, a novel upper  bound on engine's efficiency was received in terms of the fluctuations of output and input currents which is tighter than the celebrated Carnot bound \cite{bijay-universal, bijay-engine1, bijay-engine2}. A natural question that immediately arises following the work in Ref.~\cite{bijay-universal} is what happens when Onsager's reciprocity relation is broken, or in other words, the time-reversal symmetry is not respected.  Such a situation is quite common in many scenarios, for example, thermoelectric transport in presence of magnetic field \cite{keiji-strong,brandner-multi, review-engine, thermo-broken1,thermo-broken2,thermo} or cyclic heat engines driven in a time-asymmetric manner \cite{keiji-udo, cyclic}. This is the gap we fill in this paper by extending our earlier findings \cite{bijay-universal} to the broken time reversal symmetry case.

In particular, in this paper we focus on a general class of machines and show that even in the absence of time-reversal symmetry strict bounds exist between the relative fluctuations of output and the input currents in the useful operational regimes such as engine, refrigerator or pump. However, in this general case, the relative fluctuation of each current should involve the sum of  both forward and reverse currents. The result obtained here therefore generalises the previous work in Ref.~\cite{bijay-universal} significantly and can be applied beyond autonomous steady-state setups such as for periodically driven cyclic heat engines. 
%In the symmetric case, we recover the recently obtained results in Ref.~(\cite{bijay-universal}).  
We further make the connection of our study with the  generalized TUR (GTUR) relations for the time-reversal broken case. To  illustrate all these finding we consider two different paradigmatic setups : (i) a steady-state three-terminal quantum thermoelectric setup subjected to an external magnetic field, and (ii) a periodically driven classical cyclic Brownian heat engine. 

\noindent{\it Universal bounds in absence of time-reversal symmetry.--}  
We consider here a  generic out-of-equilibrium setup with two independent stochastic currents ${\cal J}_1$ and ${\cal J}_2$ that are generated by two corresponding conjugate  affinities ${\cal F}_1$ and ${\cal F}_2$, respectively. In the linear response (LR) regime, these currents are expanded upto linear order in affinities and expressed in terms of the Onsager's kinetic coefficients ${\cal L}_{ij}$ as  $\la {\cal J}_i \ra = \sum_{j=1,2} {\cal L}_{i j} {\cal F}_j, \,\, i=1,2$ \cite{Onsager1,Onsager2}. In general
${\cal  L}_{12} \neq {\cal L}_{21}$ corresponds to a broken time-reversal situation. These kinetic coefficients satisfy the universal bounds \cite{review-engine, thermo-broken1,thermo-broken2,eff-stat-bijay}
\be
{\cal L}_{11}, {\cal L}_{22} \geq 0, \quad \quad {\cal  L}_{11} {\cal  L}_{22} - \frac{\big({\cal  L}_{12} + {\cal  L}_{21}\big)^2}{4} \geq 0
\label{Onsager-bound}
\ee
which are obtained from the second law of thermodynamics that ensures the non-negativity of the net entropy production rate $\langle \sigma \rangle = \sum_{i=1,2} \la {\cal J}_i \ra \, {\cal F}_i = \sum_{i, j=1,2} {\cal L}_{ij} {\cal F}_i {\cal F}_j\geq 0$ \cite{review-engine, st-thermo1}.

Given the above process, which is from now onwards referred to as the time-forward (F) process, we consider next a corresponding time-reversed (R) process. As an example, in case of an autonomous steady-state transport setup, a time-reversed process is realized by reversing the sign of the external magnetic field. We express the currents in the R process as  $\bar{\la {\cal J}_i \ra} = \sum_{j=1,2} \bar{ {\cal L}}_{i j} {\cal F}_j$ where $\bar{{\cal L}}_{ij}$ are the kinetic coefficients for the $R$ process and can be related to the coefficients in the $F$ process via the Onsager-Casimir relation  ${\cal L}_{ij} = \bar{{\cal L}}_{ji}$ \cite{review-engine, thermo-broken1,thermo-broken2,eff-stat-bijay}.  Note that, we assume the thermodynamic affinities ${\cal F}_1, {\cal F}_2$ here to be time-reversal symmetric. Throughout the paper, we follow the convention that currents flowing into the system are positive.
Considering now the fluctuations of currents (denoted by double angular brackets) $\la \la {\cal J}^2_i \ra \ra = \la {\cal J}^2_i \ra - \la {\cal J}_i \ra^2$, for both F and R processes, we define the squared relative uncertainty for individual current $i$ as, 
\be 
\epsilon^2_i = \frac{ \langled {\cal J}_i^2 \rangled + \bar{ \langled {\cal J}_i^2 \rangled }} { \big (\langle {\cal J}_i \rangle+ \bar{\langle {\cal J}_i\rangle} \big)^2 }.
\ee
Next, following a similar procedure as in Ref.\cite{bijay-universal}, we construct the ratio between the relative uncertainties of the two currents which in the LR regime can be expressed solely in terms of the symmetric components of the Onsager coefficients of the F process, thanks to the Onsager-Casimir symmetry. We receive 
\be
{\cal Q} \equiv \frac{\epsilon^2_2}{\epsilon^2_1} = \frac{{\cal  L}^{s}_{22}}{{\cal  L}^{s}_{11}} \frac{\sum_{ij} {\cal  L}^{s}_{1i} {\cal  L}^{s}_{1j} {\cal F}_i {\cal F}_j} {\sum_{ij} {\cal  L}^{s}_{2i} {\cal  L}^{s}_{2j} {\cal F}_i {\cal F}_j}.
\label{eq:def}
\ee
 Here the symmetric components are defined as  ${\cal L}_{ij}^{s} = ({\cal L}_{ij} + {\cal L}_{ji})/2=({\cal L}_{ij} + \bar{{\cal L}}_{ij})/2$. Note that in the LR regime we substitute the current fluctuations by their corresponding  kinetic coefficients i.e., 
$\langled {\cal J}^2_{i}\rangled_{\rm eq}= \langled {\bar{\cal J}}^2_{i}\rangled_{\rm eq}=2\,{\cal L}^{s}_{ii}$, which is  the standard fluctuation-dissipation relation in equilibrium \cite{FDT3}.
One can further manipulate the above expression and re-write 
\be
{\cal Q} - 1= \frac{4}{{\cal  L}^{s}_{11}  \big (\langle {\cal J}_2 \rangle \!+\! \bar{\langle {\cal J}_2\rangle} \big)^2}  \sum_{ij} \Big[{\cal  L}^{s}_{22}  {\cal  L}^{s}_{1 i} {\cal  L}^{s}_{1 j} -  {\cal  L}^{s}_{11} {\cal  L}^{s}_{2 i}  {\cal  L}^{s}_{2 j} \Big] {\cal F}_{i} {\cal F}_{j}.
\ee
The above summation can be expanded and the finite contribution comes only from $i=j$ terms. As a result, we receive
\be
{\cal Q} -1 \!= \frac{4}{{\cal  L}^{s}_{11}  \Big (\langle {\cal J}_2 \rangle \!+\! \bar{\langle {\cal J}_2\rangle} \Big)^2} \! \Big[{\cal  L}^{s}_{11} {\cal F}_1^2 - {\cal  L}^{s}_{22} {\cal F}_2^2\Big]\, \Big[{\cal  L}^{s}_{11} {\cal  L}^{s}_{22} \!-\! ({\cal  L}^{s}_{12})^2\Big].
\label{Eq:Qbound2}
\ee
Interestingly, the last term in the above expression is the determinant of the symmetric part of the Onsager's matrix which is always non-negative and directly follows from Eq.~(\ref{Onsager-bound}). So-far, the analysis has been general with no condition imposed on the direction of the currents flowing across the system. However, to realise a useful operational machine such as an engine or a refrigerator or a pump, we need to identify the directions of input and output currents. To do this, from here onwards, we identify $\la {\cal J}_1\ra$ and $\la {\cal J}_2 \ra$ as the input and the output currents, respectively, in the F process and similarly $\la \bar{{\cal J}}_1\ra$ and $\la \bar{{\cal J}_2}\ra$ as the input and output currents in the R process, respectively.  First focusing on the input current in the F (R) process, we demand that
$\langle {\cal J}_1\rangle {\cal F}_1 > 0$, ( $\bar{\langle {\cal J}_1\rangle} {\cal F}_1 > 0$),  which in the LR regime generates a condition involving the affinities and the kinetic coefficients as ${\cal  L}_{11} {\cal F}_1^2 > - {\cal  L}_{12} {\cal F}_1 {\cal F}_2 $ ( ${\cal  L}_{11} {\cal F}_1^2 > - {\cal  L}_{21} {\cal F}_1 {\cal F}_2 $).  A combination of these two conditions yields
\be 
{\cal  L}^{s}_{11} {\cal F}_1^2 > -{\cal  L}^{s}_{12} {\cal F}_1 {\cal F}_2. 
\label{eq:in1}
\ee
We next demand that in the operational regime, the setup delivers output in both the F and R processes and thus 
$- \langle {\cal J}_2 \rangle {\cal F}_2 > 0$, ( $-\bar{\langle {\cal J}_2\rangle} {\cal F}_2 > 0$) which translates to another condition in the LR regime,
\be 
-{\cal  L}_{12}^{s} {\cal F}_1 {\cal F}_2 > {\cal  L}^{s}_{22} \, {\cal F}_2^2.
\label{eq:in2}
\ee
 Combining Eqs.~(\ref{eq:in1}) and (\ref{eq:in2}), we receive an important inequality 
 \be
 {\cal  L}^{s}_{11} {\cal F}_1^2 - {\cal  L}^{s}_{22} {\cal F}_2^2 > 0.
 \label{central-ineq}
 \ee
This is exactly one of the terms that appears in Eq.~(\ref{Eq:Qbound2}). As a result of this inequality and Eq.~(\ref{Onsager-bound}), we arrive at an important conclusion that for a generic out-of-equilibrium setup, operating as a useful device in both the F and the R processes, ${\cal Q} \geq 1$.  This is the one of central results of this paper. Because of the strict inequality in Eq.(\ref{central-ineq}), ${\cal Q}=1$ in Eq.~(\ref{Eq:Qbound2}) can only be reached when the determinant of the symmetric part of the Onsager's matrix vanishes. This corresponds to a situation when $\big(\la {\cal J}_1 \ra + \la \bar{\cal J}_1 \ra\big) \propto \big(\la {\cal J}_2 \ra + \la \bar{\cal J}_2 \ra\big)$, which is a generalized version of the so-called tight-coupling limit as it involves both forward and reversed currents. For time-reversal symmetric systems this generalized coupling corresponds to the usual tight coupling limit i.e.,  $\la {\cal J}_1 \ra  \propto \la {\cal J}_2 \ra$\cite{bijay-universal}. Interestingly, $\la {\cal J}_1 \ra  \propto \la {\cal J}_2 \ra$ can never be attained in the broken time reversal situation as it violates the standard second law of thermodynamics (please see Appendix A for the details).   A few important outcomes follow immediately from the above central result. To proceed further, we first define the ratio between fluctuations of output and input currents, scaled by the affinities, and express it in the LR regime as,
\be
\eta^{(2)} \equiv \frac{ {\cal F}_2^2 \, \langled {\cal J}_2^2 \rangled}{{\cal F}_1^2 \, \langled  {\cal J}_1^2 \rangled} = \frac{ {\cal F}_2^2 \, L^{s}_{22} }{{\cal F}_1^2 \,   L^{s}_{11}}.
\ee
Note that in the LR regime $\eta^{(2)}$ is identical to the reversed $\bar{\eta}^{(2)}$, defined using the fluctuations in R process. Now, for the time-reversal symmetric (TRS) case, the F and the R processes are identical, i.e., ${\cal  L}_{12}={\cal  L}_{21}$ and following the bound ${\cal Q} \geq 1$,  $\eta_{\rm TRS}^{(2)}$ receives an universal lower bound given by the square of the mean efficiency, 
\be
\eta_{\rm TRS}^{(2)} \geq \Big[\frac{-{\cal F}_2 \, \langle {\cal J}_2 \rangle}{{\cal F}_1 \, \langle {\cal J}_1 \rangle}\Big]^2 = \langle \eta \rangle_{\rm TRS}^2.
\label{lower}
\ee
where $\la \eta \ra = \frac{- \la {\cal J}_2 \ra {\cal F}_2}{\la {\cal J}_1 \ra {\cal F}_1}$ is the scaled mean thermodynamic efficiency. 
This result matches with the one obtained in Ref.~\cite{bijay-universal}. In the general case, ${\cal  L}_{12} \neq {\cal  L}_{21}$, however, the lower bound  for $\eta^{(2)} \geq \la \eta\ra ^2 $ is not respected for F or R processes independently.  One can however restore a similar looking bound by considering a generalized mean efficiency $\la \eta \ra_{G}$ defined as the ratio of sum of the output currents in both F and R processes to the corresponding sum of input currents, i.e., 
\be
\la \eta \ra_{G} = \frac{-{\cal F}_2 \,\big( \langle {\cal J}_2 \rangle + \langle \bar{\cal J}_2 \rangle\big)}{{\cal F}_1 \, \big(\langle {\cal J}_1 \rangle + \langle \bar{\cal J}_1 \rangle\big)} 
\ee
which following ${\cal Q} \geq 1$ then gives a universal lower bound,
\be
\eta^{(2)} =  {\bar{\eta}}^{(2)} \geq \la \eta \ra^{2}_{G}.
\label{uni-lower}
\ee
Furthermore, for $\eta^{(2)}$ an universal upper bound also exists due to the inequality in Eq.~(\ref{central-ineq}), employing which we receive,   
\be
\eta^{(2)}  =  {\bar{\eta}}^{(2)}  = \frac{ {\cal F}^2_2 {\cal  L}^s_{22}} {{\cal F}^2_1 {\cal  L}^s_{11}}  < 1.
\label{uni-upper}
\ee
Interestingly, the same upper bound was received for continuous engines \cite{bijay-universal} in time reversal symmetric case.  Our result therefore implies that the upper bound for $\eta^{(2)}$ is robust even in the absence of time-reversal symmetry. Note that, in the context of thermal engines $(\rm ENG)$ with a working fluid operating between a hot and a cold reservoir with fixed temperatures $T_h$ and $T_c$, respectively, taking into account the proper definitions for input heat and output work, the above scaled upper bound translates to 
\be
 \eta_{\rm ENG}^{(2)} = \bar{\eta}_{\rm ENG}^{(2)}  < \eta_C^2.
\ee
This result can be easily extended to other operational regimes like refrigerator or pump. 

As a final point, we note that our result provides a strong and  deep connection with the recently obtained generalized thermodynamic uncertainty relation (GTUR) \cite{Horowitz-trb} for time-reversal broken systems. The GTUR in the LR regime provides a universal lower bound on the relative fluctuations $\epsilon_i^2$ of individual currents in terms of the associated entropy production rate. The GTUR reads  \cite{Horowitz-trb}, 
\be
\Big(\langle \sigma \rangle + \la \bar{\sigma} \ra\Big)\, \epsilon^{2}_i  \geq 2, \quad i=1,2.
\ee
The proof of GTUR in the LR regime requires only the non-negativity condition of the symmetric part of the Onsager's matrix. Once again, the equality here corresponds to a generalized tight-coupling situation. In general, no obvious relation exists between the GTUR for different currents. However as shown in this work,  in the operational regime, $\epsilon_2^2 \geq \epsilon_1^2$, immediately implies a strict hierarchy for the GTUR between the output and the input currents,
\be
\Big(\langle \sigma \rangle + \la \bar{\sigma} \ra \Big) \, \epsilon^{2}_2  \geq \,
 \Big(\langle \sigma \rangle + \la \bar{\sigma} \ra \Big)\, \epsilon^{2}_1 \geq 2.
\label{ineq:G-TUR}
\ee
Note that, in the LR regime the entropy production rate is same for both F and R processes. Therefore the above expression further simplifies to $\la \sigma \ra  \epsilon^{2}_2 \geq \la \sigma \ra  \epsilon^{2}_1 \geq 1$. Eqs.~(\ref{uni-lower}), (\ref{uni-upper}) and (\ref{ineq:G-TUR}) are the other central results of this paper. We would like to stress that the results obtained here are universal in the LR regime and valid for both classical and quantum systems. In what follows, we illustrate these results for two paradigmatic setup with broken time-reversal symmetry. We will first consider an autonomous steady-state quantum dot setup where time-reversal symmetry is broken via an external magnetic field and as a second example we focus on a periodically driven classical cyclic Brownian heat engine.  

\begin{figure}
\centering
\mbox{\subfigure{\includegraphics[height=1.6in, width=1.9in, scale=0.5]{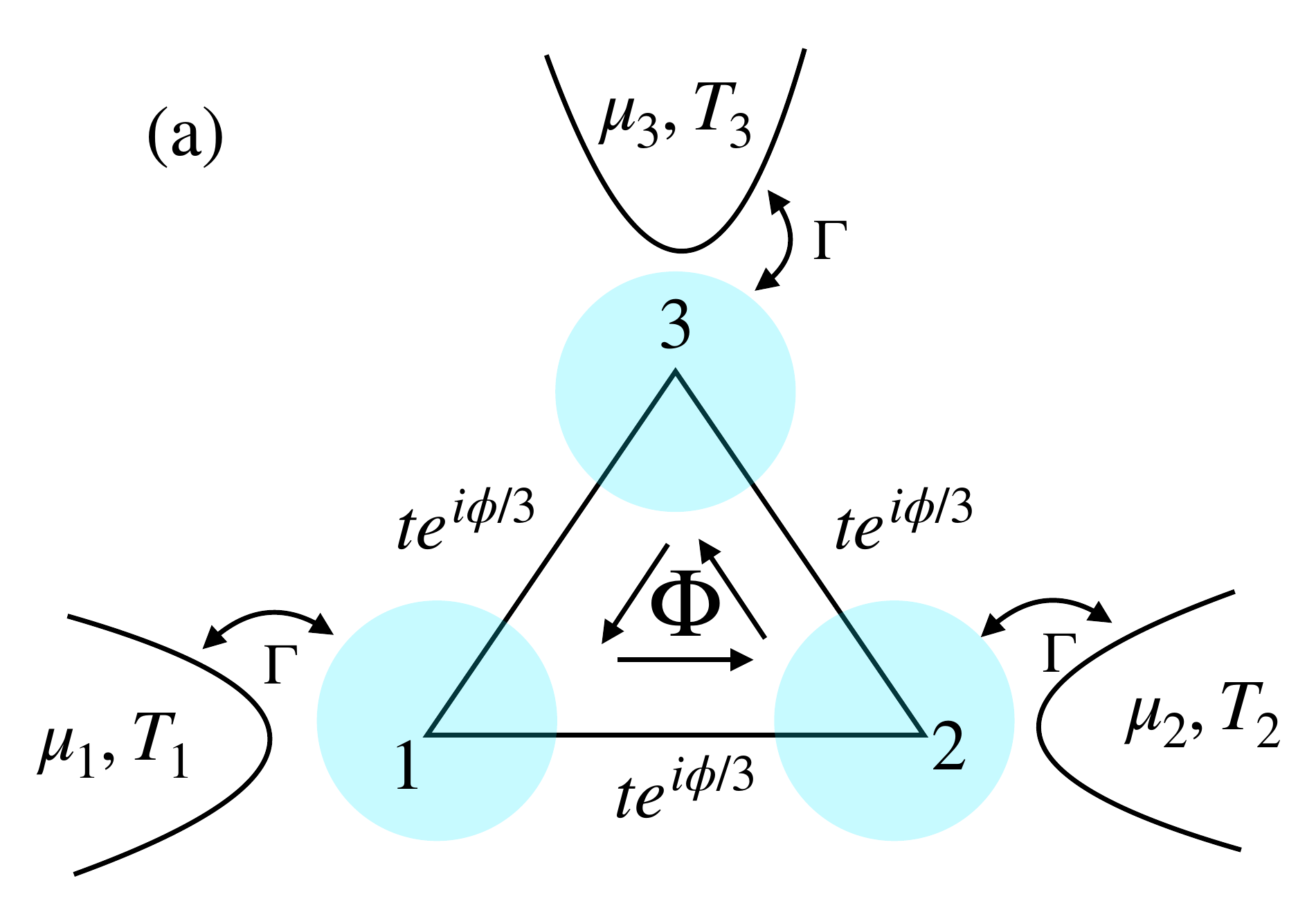}}\quad
\subfigure{\includegraphics[height=1.6in, width=1.5in, scale=0.5]{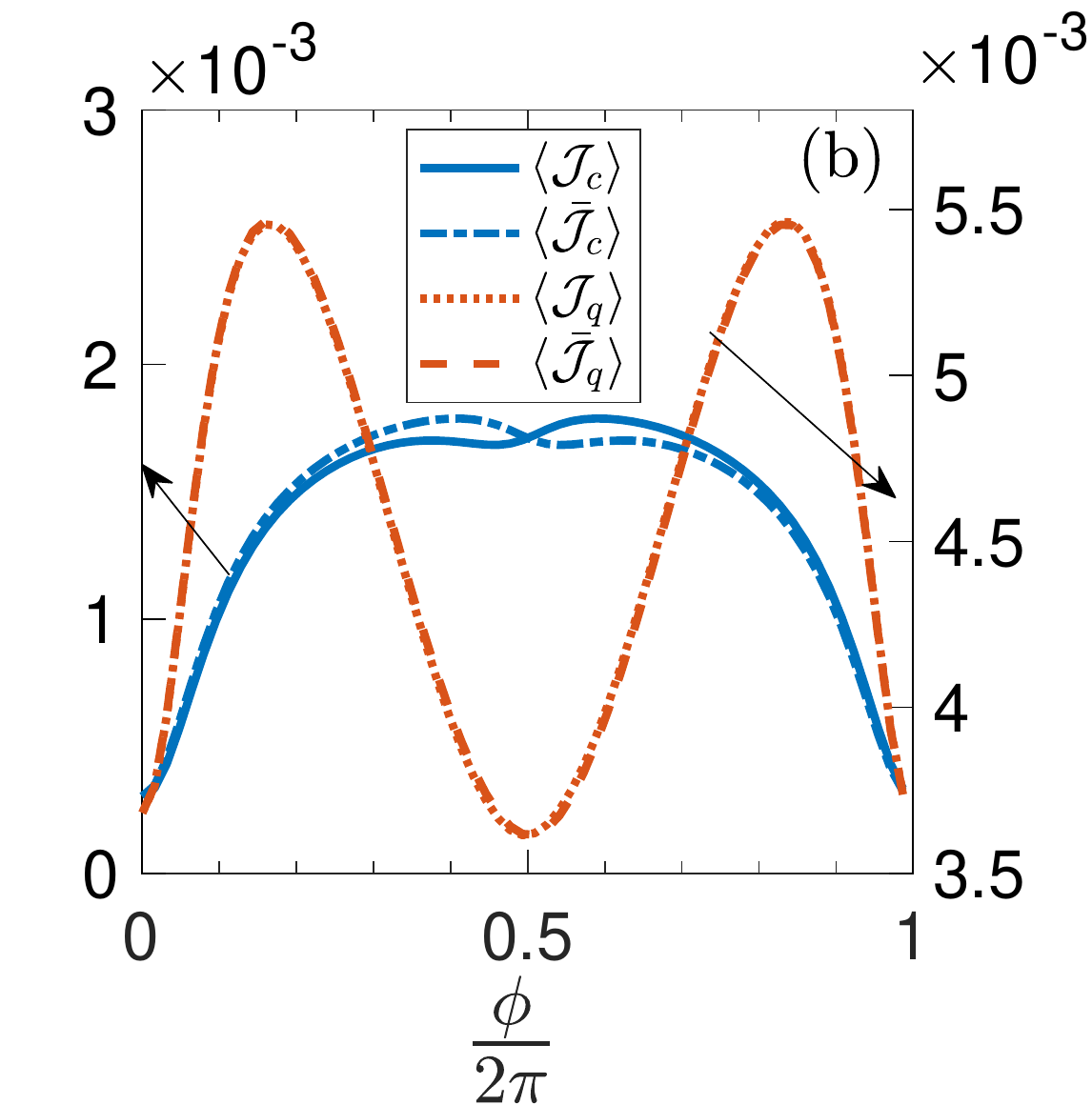} }}
\includegraphics[height=2in, width=3.5in, scale=0.5]{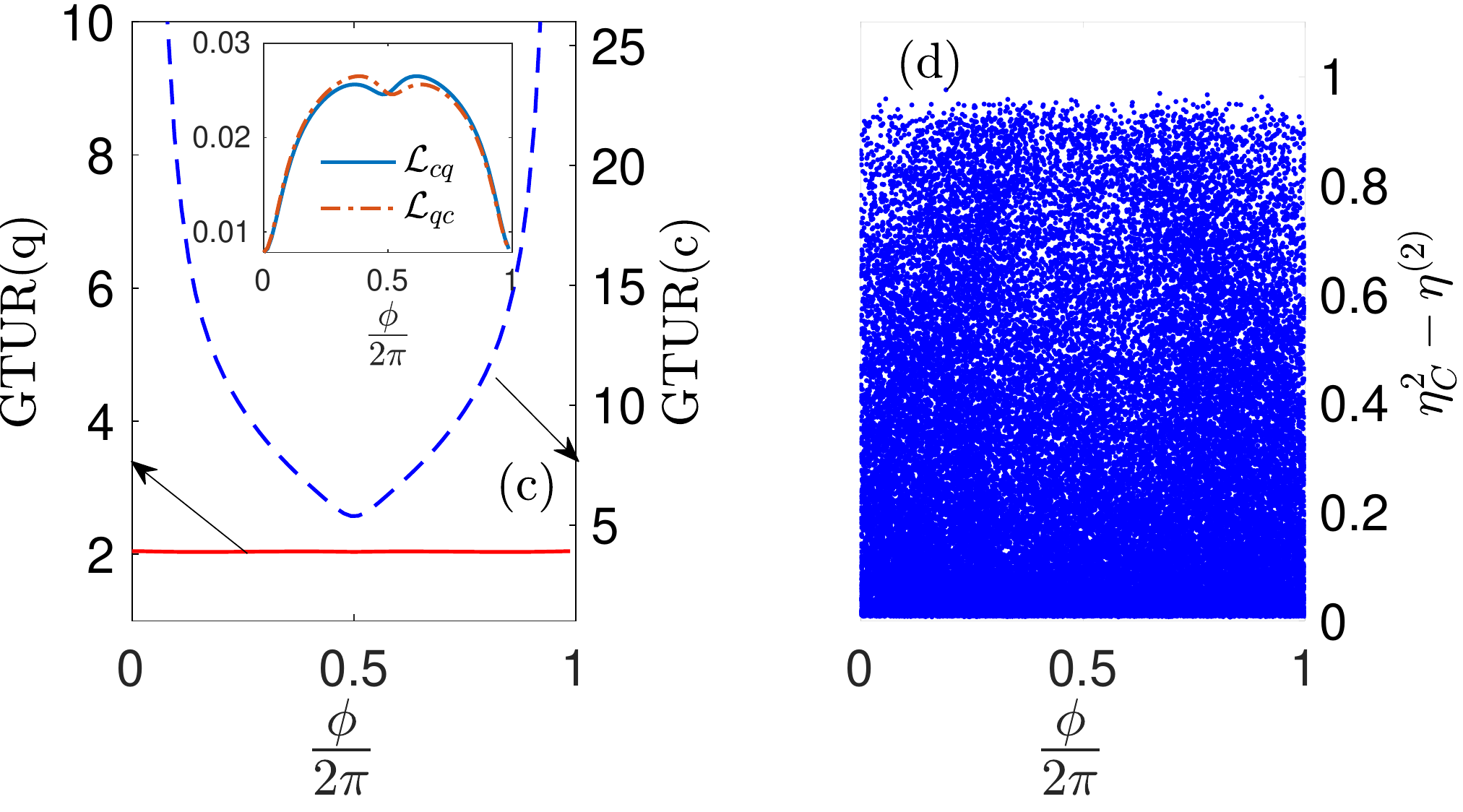}
\caption{(Color online): (a) Schematic of a thermoelectric setup consisting of three quantum dots with each dot further in contact with its own thermochemical bath. The third bath acts as a B\"uttiker probe whose temperature $T_3$ and chemical potential $\mu_3$ are fixed by imposing zero net heat and particle currents from the probe. (b) Plot for steady-state heat and charge currents flowing out of the left bath in F ( ($\la {\cal J}_q \ra, \la {\cal J}_c\ra$)) and R ($\la \bar{\cal J}_q \ra, \la \bar{\cal J}_c\ra$) processes, as a function of the phase $\phi$. As per our convention and with the bias settings  $T_1 > T_2$ and $\mu_2 > \mu_1$, the setup  works as a thermoelectric engine in both F and R processes when $\la {\cal J}_q \ra, \la {\cal J}_c\ra, \la \bar{\cal J}_q \ra, \la \bar{\cal J}_c \ra > 0$. (c) The generalized TUR (GTUR) ratio for input heat current from the left (hot) bath and the output charge current. The inset in (c) shows the off-diagonal elements of the Onsager matrix as a function of the phase $\phi$. (d) A scatter plot showing the validity of the upper bound for $\eta^{(2)}$. We display $\eta_C^2 - \eta^{(2)}$ as a function of $\phi$ over 50000 sample points  choosen randomly. For figures (b) and (c) the parameters are choosen as $\epsilon_{1}=2.0$, $\epsilon_{2}=1.3,$ $\epsilon_{3}=0.9$, $t=-1$ and $\Gamma=0.5$, $T_1=1.1, T_2= 1, \mu_1=-0.01, \mu_2=0.01$. For the plot (d), $\epsilon_{1}, \epsilon_{2}$, and $\epsilon_{3}, \Gamma, T_2$ are chosen randomly in the interval between $[0,1]$, $t$ is chosen between $[-1,0]$, $\mu_1$ is chosen from $[-0.01,0]$ and $\mu_2$ is chosen from $[0, 0.01]$.}
%\end{center}
\label{thermo1}
\end{figure}

{\it Example I: Quantum thermoelectric transport in a three-dot setup in presence of an external magnetic field:}
As a first example we consider a mesoscopic triple quantum dot (QD) thermoelectric setup which is arranged in a triangular geometry and is additionally pierced by an Aharonov-Bohm magnetic flux $\Phi$ (Fig.~\ref{thermo1}(a)).  Each dot is further in contact with its own thermochemical reservoir which is maintained in equilibrium at a fixed temperature $T_{i}$ and a chemical potential $\mu_{i}$, where $i=1, 2, 3$ corresponds to left, right, and probe terminals, respectively. The Hamiltonian of this entire setup is given by $H=H_d+H_b+H_{tun}$ where $H_d= \sum^3_{i=1} \epsilon_i  a^{\dagger}_i a_i+ t e^{i \phi/3} (a^{\dagger}_{2} a_1 + a^{\dagger}_{3} a_2+a^{\dagger}_{1} a_3)+ {\rm h.c}$ is the Hamiltonian for the three QD's. Here  $a_i (a_i^{\dagger})$ is the electronic annihilation (creation) operator for the $i$-th dot,  $\epsilon_i$ is the onsite energy, $t$ is the hopping parameter between the dots and $\phi= 2 \pi \Phi/\Phi_0$ is  the phase picked up by the electron by moving around the loop once with  $\Phi$ being the magnetic flux penetrating the setup and $\Phi_0=hc/e$ is the flux quantum.
The baths are modelled as non-interacting free electron gas with Hamiltonian $H_b=\sum_{ik} \epsilon_{ik} b^{\dagger}_{ik} b_{ik}$ 
and $H_{tun}= \sum_{ik}V_{ik} a^{\dagger}_i b_{i k} + {\rm h.c}.$ describes the tunneling of electrons between the dots and the baths.
The coupling strength between the baths and the dots is characterized by $\Gamma=\sum_{k} |V_{ik}|^2 \delta(\epsilon-\epsilon_{k})$ which is assumed to be identical and energy independent for all the three baths.
%We denote the dots with 

Given the fully non-interacting nature of the setup, the transport properties can be analytically calculated following the standard Landauer-B\"uttiker formalism \cite{keiji-strong, brandner-multi}. We are interested here to realize a thermoelectric engine that absorbs heat current (input) from the hot bath and push the electrons against the chemical bias and thereby generating electric current (output). We set $T_1 > T_2$ and $\mu_2 > \mu_1$ and fix $T_3$ and $\mu_3$ for the third bath by demanding that the net charge and heat currents flowing out of this bath is zero. The third bath therefore acts as a B\"uttiker probe that allows to simulate elastic and inelastic scattering processes within the central QD system \cite{Bu-1, Bu-2}.
As a result of this probe condition, it is sufficient for our analysis to consider currents flowing out either of the left or the right bath. We consider the right bath as reference and define the affinities for heat current ${\cal F}_q = 1/T_2 \!-\! 1/T_1>0$ and particle current ${\cal F}_c= (\mu_1 \!-\! \mu_2)/T_2<0$ and write the current flowing out of left bath as $\la {\cal J}_{\alpha}\ra= \sum_{\beta=c, q} {\cal L}_{\alpha \beta} {\cal F}_{\beta}$,  $\alpha=(c,q)$ for the F process. Analogous expression for the R process is obtained by reversing the phase $\phi \to -\phi$ in the dot Hamiltonian $H_d$. The presence of the magnetic field and the B\"uttiker probe ensures ${\cal L}_{c q}(\phi) \neq {\cal L}_{qc}(\phi)$, in general. It can be easily checked numerically that the Onsager-Casimir relation remains intact i.e., ${\cal L}_{c q} (\phi) = {\cal L}_{qc} (-\phi)$, as expected.
 
In Fig.~(\ref{thermo1}(b)) we display the forward ($\la {\cal J}_q \ra, \la {\cal J}_c\ra$) and the backward currents ($\la \bar{\cal J}_q \ra, \la \bar{\cal J}_c\ra$) as a function of $\phi$, fulfilling the engine condition ($\la {\cal J}_q \ra \geq 0$, $\la {\cal J}_c\ra \geq 0$ ) in F and ($\la \bar{\cal J}_q \ra \geq 0$, $\la \bar{\cal J}_c \ra \geq 0$)  R processes.  Fig.~(\ref{thermo1}(c)) shows the validity of one of our central results (Eq.~(\ref{ineq:G-TUR})) with GTUR product $\Big((\langle \sigma \rangle + \la \bar{\sigma} \ra) \, \epsilon^{2}_c\Big)$  for the output charge current is always larger than the corresponding GTUR product of the input heat current $\Big((\langle \sigma \rangle + \la \bar{\sigma} \ra) \, \epsilon^{2}_q\Big)$ from the hot bath in the engine regime. The inset display the off-diagonal Onsager coefficients, indicating the broken time-reversal symmetry due to the presence of magnetic field and the B\"uttiker probe. In Fig.~(\ref{thermo1}(d)) we show the validity of the universal upper bound for $\eta_{\rm ENG}^{(2)} \equiv \frac{ \la \la {\cal J}^2_w \ra \ra}{ \la \la {\cal J}^2_q \ra \ra} <\eta_C^2$ by choosing the parameters of the model randomly.  For the thermoelectric setup, the work fluctuation is identified as $\la \la {\cal J}^2_w \ra \ra = (\mu_2\!-\!\mu_1)^2 \la \la {\cal J}^2_c \ra \ra$.

Interestingly, for this setup a stronger constraint on Onsager coefficients was obtained in Ref.~\cite{keiji-strong} which is given as 
\be
{\cal L}_{cc} {\cal L}_{qq} - \big({\cal L}_{cq} + {\cal L}_{qc}\big)^2 /4 \geq \frac {3}{4} \big({\cal L}_{cq} -{\cal L}_{qc}\big)^2 \geq 0.
\ee
As a consequence, the determinant of the symmetric part of the Onsager's matrix will never be zero whenever ${\cal L}_{cq} \neq {\cal L}_{qc}$ and thus ${\cal Q}$ can never attain the value 1 for this setup, except when ${\cal L}_{cq} = {\cal L}_{qc}$.

{\it Example II:  Classical periodically driven cyclic Brownian heat engine:}
As a second example we consider a one-dimensional classical stochastic cyclic Brownian heat engine. This setup consists of an overdamped Brownian particle trapped in a harmonic potential with stiffness $\kappa(t)$ and bath temperature $T(t)$, both varying periodically in time. 
This particular model was first proposed in Ref.~\cite{brownian-theory} and later experimentally realized in Ref.~\cite{ brownian-expt}. The Hamiltonian for the particle in the overdamped regime is written as 
$H(x,t)=
\frac{\kappa_0}{2}x^2+\Delta H g_w(x,t)$ where $\Delta H =\kappa x_0^2$ is the strength of the mechanical perturbation with $x_0=\sqrt{2T/\kappa_0}$ being the characteristic length scale of the model and $g_w(x,t)=
%g_w(x)\gamma_w(t)=
\frac{x^2}{2x_0^2}\gamma_w(t)$. The Brownian particle is also subjected to a periodically changing temperature profile, given as, $T(t)=T+\Delta T\gamma_q(t)$.  Both $\gamma_w(t)$ and $\gamma_q(t)$ here are arbitrary ${\cal T}$-periodic functions. The probability distribution $p(x,t)$ of finding the particle at the position $x$ at time $t$ evolves as per the  Fokker-Planck equation $\partial_t p(x,t)= L(t) p(x,t)$ with the Fokker-Planck operator given as \cite{Risken} 
\be
L(t)=\frac{1}{\gamma} \big[\kappa_0 + k \gamma_w(t)\big] \partial_x x+\frac{T(t)}{\gamma}\partial_x^2
\ee
where $\gamma$ is the friction coefficient. We closely follow here the prescription of Kay Brandner et. al \cite{keiji-udo} where a linear response analysis has recently been carried out for such setup by considering $\Delta T$ and $\Delta H$ as small parameters. In the periodic steady state, it turns out the net entropy production rate, averaged over one cycle, can be expressed in the standard form $\la \sigma \ra= \sum_{\alpha=w, q} {\cal F}_{\alpha} \, \la {\cal J}_{\alpha}\ra$ with $\la {\cal J}_{\alpha}\ra=\sum_{\beta=w, q} {\cal L}_{\alpha \beta }{\cal F}_{\beta}, \alpha=w,q$ being the current and the affinities for work and heat currents are given as 
${\cal F}_w=\Delta H /T$ and ${\cal F}_q=\Delta T/T^2$, respectively, which are considered to be positive here.  The expressions for the kinetic coefficients ${\cal L}_{\alpha \beta}$ can be obtained by solving the Fokker-Planck equation following a first-order perturbative expansion in $\Delta H$ and $\Delta T$ of $L(t)$. 

For a clear demonstration, we choose specific protocols $\gamma_w(t)= \sin{\big(2\pi t/{\cal T}+\phi\big)}$ and $\gamma_q(t)=\big(1+\sin{(2\pi t/{\cal T})}\big)/2$
for which explicit expressions for the Onsager coefficients can be obtained \cite{Garrahan18} (please see Appendix B for the details), 
\begin{equation}
    {\cal L}=\frac{\lambda(1-\chi)}{16}
    \begin{bmatrix}
    1 & -T(\cos{\phi}+\frac{\lambda}{\Omega}\sin{\phi}) \\
    -T(\cos{\phi}-\frac{\lambda}{\Omega}\sin{\phi}) & T^2
    \end{bmatrix}
    \label{ons-mat}
\end{equation}
where $\lambda=2\kappa_0/\gamma$ , $\Omega=2\pi/{\cal T}$ and $\chi=\lambda^2/(\lambda^2+\Omega^2)$. Thus, $0\le\chi\le 1$ with $\chi\to1$ characterizing the quasi-static limit with vanishing entropy production. 
%In this limit of large cycle time, we see that all the currents also vanish. This is intuitively expected because here the currents are defined over one complete cycle.
For the chosen protocols $\gamma_w(t), \gamma_q(t)$, the R process is realized either by substituting $t \to -t$ or $\phi \to -\phi$.  The final effect on the Onsager matrix is simply to swap the off-diagonal elements. 

\begin{figure}
\centering
\includegraphics[trim=0 0 0 0, clip=Flase, width=\columnwidth]{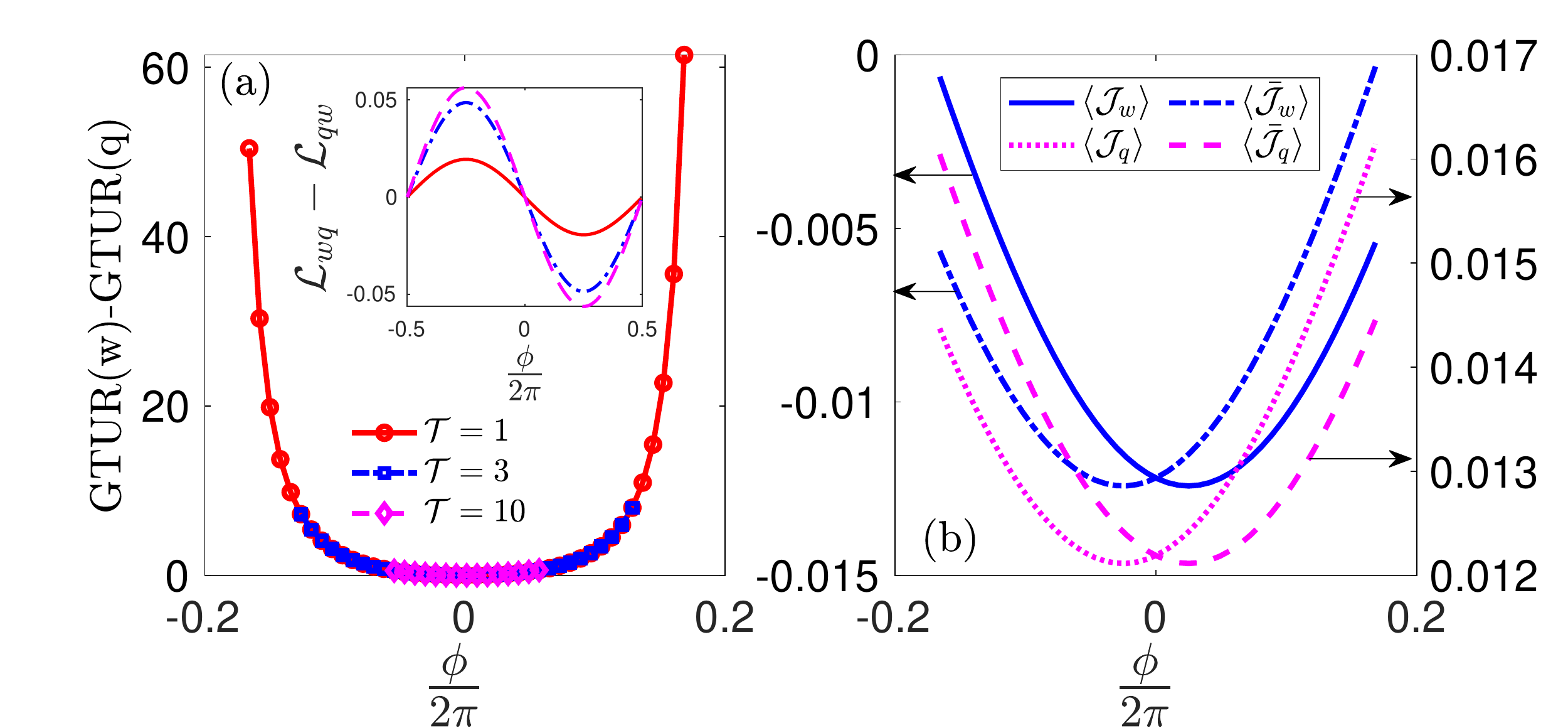}
\caption{(Color online:) Results for a classical cyclic Brownian heat engine: (a) Plot for the difference between GTUR for output work current (GTUR(w)) and GTUR for input heat current (GTUR(q)) in the engine regime as a function of $\phi$ for three different cycle times ${\cal T}=1, 3$, and $10$. The inset shows the difference between the off-diagonal elements of Onsager matrix (${\cal L}_{wq} \!-\! {\cal L}_{qw}$) for the same cycle times. (b) Plot for time-averaged work and heat currents in the engine regime for both F and R processes for ${\cal T}=1$. As both $\Delta H$ and $\Delta T$ are chosen positive here, following our convention an engine is realized when $\la {\cal J}_w \ra, \la\bar {\cal J}_w \ra<0$, and $\la {\cal J}_q \ra, \la\bar {\cal J}_q \ra>0$. The parameters considered here are $T=1, k_0=1, \gamma=2, \Delta T=0.3, \Delta H=0.1$.}
\label{CHE}
\end{figure}
We demonstrate the results in Fig.~(\ref{CHE}) by focusing once again on the engine regime and display the GTUR bounds (Fig.~(\ref{CHE}) (a)) as a function of $\phi$ for three different cycle times. Fig.~(\ref{CHE}) (b) shows the currents (shown only for ${\cal T}=1$) satisfying the engine conditions in both F and R processes i.e., $\la {\cal J}_w\ra, \la \bar{\cal J}_w\ra <0, \la {\cal J}_q \ra,  \la \bar{\cal J}_q \ra>0$, for positive  $\Delta H$ and $\Delta T$. As expected, the GTUR for output work current is always upper bounded by the corresponding GTUR for input heat current with the difference (GTUR(w)- GTUR(q)) gets reduced with increasing cycle time ${\cal T}$, corresponds to approaching the generalized tight-coupling limit. Following the positivity of the entropy production rate, it is easy to check that the standard thermodynamic efficiency for the engine, defined as, $\la \eta \ra_{\rm ENG}=-\frac{T{\cal F}_w \, \la {\cal J}_w \ra }{\la {\cal J}_q\ra} \leq \eta_C= \frac{\Delta T}{T}$ is bounded by the Carnot value. Similarly, following Eq.~(\ref{ons-mat}), we receive for $\eta^{(2)}_{\rm ENG}$,
\be
\eta_{\rm ENG}^{(2)}\equiv \frac{T^2 \, {\cal F}^2_w \la \la {\cal J}^2_w \ra \ra }{\la \la {\cal J}^2_q \ra \ra}= \Big(\frac{\Delta H}{T}\Big)^2
\ee
which upon imposing the engine condition (Eq.~(\ref{central-ineq})) generates the condition $(\Delta H)^2 < (\Delta T)^2$ and we thus receive, in the engine regime the upper bound, $\eta^{(2)}_{\rm ENG} < \eta_C^2$.

In summary, we have generalised the previous study on universal bounds on fluctuations for machines in a significant way by incorporating time-reversal symmetry breaking situation. We show that even in this general situation non-trivial universal upper and lower bounds for $\eta^{(2)}$ exist whenever a setup operates as a useful machine. However in order to receive such bounds the relative fluctuations of the sum of forward and reversed currents must be taken into account. As a consequence of the lower bound (${\cal Q} \geq 1$), we further able to establish the hierarchy in the GTUR bounds between the output and input currents.  Future work will be directed towards analysing the validity of these bounds beyond the linear response regime.

%===================================================================================================
BKA acknowledges the MATRICS grant MTR/2020/000472 from SERB, Government of India. BKA also thanks the Shastri Indo-Canadian Institute for providing financial support for this research work in the form of a Shastri Institutional Collaborative Research Grant (SICRG). SS acknowledges financial support from the Council of Scientific \& Industrial Research (CSIR), India (Grant Number 1061651988). SM acknowledges financial support from the CSIR, India (File number: 09/936(0273)/2019-EMR-I).
%=======================================

%==========================================================================================
\vspace{5mm}
\renewcommand{\theequation}{A\arabic{equation}}
\renewcommand{\thefigure}{A\arabic{figure}}
\renewcommand{\thesection}{A\arabic{section}}
\setcounter{equation}{0}  % reset counter
%========================================

%=================================================================

\section{Appendix A:Unattainability of tight coupling limit in broken time-reversal systems.}
In the standard tight-coupling limit $\la {\cal J}_1 \ra  = \alpha \, \la {\cal J}_2 \ra$ where $\alpha$ is a proportionality constant. This condition implies following relations among the Onsager's kinetic coefficients,
\begin{equation}
{\cal L}_{11}=\alpha \, {\cal L}_{21} \quad , \quad  {\cal L}_{12}=\alpha \, {\cal L}_{22}
\label{tc1}
\end{equation}
Given these relations  determinant of the symmetric part of the Onsager's matrix (${\cal L}^s$) reduces to,
\begin{eqnarray}
{\rm det}[{\cal L}^s] &=& {\cal  L}_{11} {\cal  L}_{22} - \frac{\big({\cal  L}_{12} + {\cal  L}_{21}\big)^2}{4}  \nonumber\\
=&& \alpha \, {\cal L}_{21} \frac{{\cal L}_{12}}{\alpha}- \frac{\big({\cal  L}_{12} + {\cal  L}_{21}\big)^2}{4}  \nonumber \\
=&& \frac{\big(4 {\cal  L}_{12} {\cal  L}_{21}-{\cal  L}_{12}^2 + {\cal  L}_{21}^2 - 2 {\cal  L}_{12} {\cal  L}_{21} \big)}{4} \nonumber \\
=&& - \frac{\big({\cal  L}_{12} - {\cal  L}_{21}\big)^2}{4}  \nonumber \\
\le&& 0
\label{tc2}
\end{eqnarray}
where the equality is achieved only when ${\cal  L}_{12} = {\cal  L}_{21}$ which corresponds to a time-reversal symmetric situation. For broken time-reversal case, in general, this violates the standard second law of thermodynamics that always requires ${\rm det}[{\cal L}^s] \ge 0$. Hence tight coupling limit can never be achieved in broken time-reversal systems.

\section*{Appendix B: Kinetic coefficients for cyclic Brownian heat engine}
In the main text we have provided the final expression for the Onsager matrix for a specific choice of protocols for $\gamma_q(t)$ and $\gamma_w(t)$. 
The different elements of the Onsager matrix for cyclic heat engine can be calculated using the formula 
\begin{widetext}
\begin{equation}
    \begin{aligned}
 \!{\cal L}_{\alpha \beta}\!=\!-\frac{2T^2{\cal E}_{\alpha} {\cal E}_{\beta}}{\cal T} \! \int_0^{\cal T}dt\Big(\dot{\gamma}_{\alpha}(t)\gamma_{\beta}(t) \!-\!\int_0^\infty d\tau\dot{\gamma}_{\alpha}(t)\dot{\gamma}_{\beta}(t\!-\!\tau)e^{-2\kappa_0\tau/\gamma}\Big)
    \end{aligned}
  \label{onsager-cyclic}
\end{equation}
\end{widetext}
where $\alpha, \beta= w,q$. For the chosen protocol in the main text it is easy to verify that a reversed dynamics corresponds to either $t \to -t$ or $\phi \to -\phi$.  Here  ${\cal E}_w=1/(4T)$ and ${\cal E}_q=-1/2$. The details about arriving to Eq.~(\ref{onsager-cyclic}) can be found in Ref.~\cite{Garrahan18,keiji-udo}.

\end{document}